\documentclass[preprint,superscriptaddress,showpacs,floatfix,prc]{revtex4}
\usepackage{amsmath}
\usepackage{graphicx}

\begin{document}

\title{Effects of momentum-dependent nuclear potential on two-nucleon
correlation functions and light cluster production in intermediate energy
heavy-ion collisions}
\author{Lie-Wen Chen}
\thanks{On leave from Department of Physics, Shanghai Jiao Tong University,
Shanghai 200030, China}
\affiliation{Cyclotron Institute and Physics Department, Texas A\&M University, College
Station, Texas 77843-3366}
\author{Che Ming Ko}
\affiliation{Cyclotron Institute and Physics Department, Texas A\&M University, College
Station, Texas 77843-3366}
\author{Bao-An Li}
\affiliation{Department of Chemistry and Physics, P.O. Box 419, Arkansas State
University, State University, Arkansas 72467-0419}
\date{\today }

\begin{abstract}
Using an isospin- and momentum-dependent transport model, we study the
effects due to the momentum dependence of isoscalar nuclear potential as
well as that of symmetry potential on two-nucleon correlation functions and
light cluster production in intermediate energy heavy-ion collisions induced
by neutron-rich nuclei. It is found that both observables are affected
significantly by the momentum dependence of nuclear potential, leading to a
reduction of their sensitivity to the stiffness of nuclear symmetry energy.
However, the t/$^{3}$He ratio remains a sensitive probe of the density
dependence of nuclear symmetry energy.
\end{abstract}

\pacs{25.70.-z, 25.70.Pq., 21.30.Fe., 24.10.Lx}
\maketitle

\section{Introduction}

The equation of state (\textrm{EOS}) of an asymmetric nuclear matter with
unequal numbers of protons and neutrons depends crucially on the nuclear
symmetry energy. Although the nuclear symmetry energy at normal nuclear
matter density is known to be around $30$ \textrm{MeV} from the empirical
liquid-drop mass formula \cite{myers,pomorski}, its values at other
densities are poorly known. Studies based on various theoretical models also
give widely different predictions \cite{ibook}. Lack of this knowledge has
hampered our understanding of both the structure of radioactive nuclei \cite%
{oya,brown,hor01,furn02} and many important issues in nuclear astrophysics %
\cite{bethe,bom,lat01}, such as the nucleosynthesis during the pre-supernova
evolution of massive stars and the properties of neutron stars. However,
recent advance in radioactive nuclear beam facilities provides a unique
opportunity to study the density dependence of nuclear symmetry energy \cite%
{ibook,npa01,ireview98,li04prc,li04npa}. Theoretical studies have already
shown that in heavy-ion collisions induced by neutron-rich nuclei, the
effect of nuclear symmetry energy can be studied via the ratio of
pre-equilibrium neutrons and protons \cite{li97}, the isospin fractionation %
\cite{fra1,fra2,xu00,tan01,bar02}, the isoscaling in multifragmentation \cite%
{betty}, the proton differential elliptic flow \cite{lis}, the
neutron-proton differential transverse flow \cite{li00,Greco03}, the $\pi
^{-}$ to $\pi ^{+}$ ratio \cite{li02}, and isospin diffusion in heavy-ion
collisions \cite{shi03,tsang04}. Also, it was found that the correlation
function between nucleon pairs with high total momentum and the isobaric
ratio t/$^{3}$He in heavy-ion collisions induced by neutron-rich nuclei are
sensitive to the density dependence of nuclear symmetry energy \cite%
{CorShort,CorLong,ClstShort,ClstLong}, implying that the space-time
properties of neutron and proton emission sources are affected by the
nuclear symmetry energy. In these studies, the momentum dependence of
nuclear mean-field potential, especially its isovector part (the symmetry
potential), was not taken into account. The momentum dependence of nuclear
isoscalar potential is well-known, and its effect in heavy-ion collisions is
large \cite{pawel02}. As shown in Refs.\cite{gbd87,gale90,pan93,zhang94},
the experimental data on the nucleon directed flow in intermediate energy
heavy-ion collisions, which were previously explained by a stiff nuclear
equation of state with compressibility of 380 MeV when the momentum
dependence of nuclear isoscalar potential was not taken into account, are
actually consistent with a soft nuclear equation of state with
compressibility of 200 MeV after including the momentum dependence. The
momentum dependence of nuclear isoscalar potential was also found to affect
the space-time properties of nucleon emission source \cite{greco99}. The
momentum dependence of nuclear symmetry potential is, on the other hand,
poorly known. Its effect on some of the isospin observables mentioned above
has only recently been studied \cite{li04a,li04b,rizzo04} and was found to
be significant. In this paper, we shall study how the momentum dependence of
both isoscalar nuclear potential and nuclear symmetry potential affect
two-nucleon correlation functions and light cluster production in
intermediate energy heavy-ion collisions induced by neutron-rich nuclei. Our
results show that the momentum dependence of nuclear potential reduces the
sensitivity of two-nucleon correlation functions and the light cluster yield
on the stiffness of nuclear symmetry energy. However, the t/$^{3}$He ratio
is still found to depend on the stiffness of nuclear symmetry energy even
with the inclusion of its momentum dependence.

This paper is organized as follows. In Section \ref{momentum}, we discuss
the momentum dependence of nuclear isoscalar potential and nuclear symmetry
potential. In Section \ref{correlation}, we present results from the
isospin- and momentum-dependent Boltzmann-Uehling-Uhlenbeck (IBUU) transport
model on two-nucleon correlation functions in heavy-ion collisions induced
by neutron-rich nuclei at intermediate energies. Results on light cluster
production and the t/$^{3}$He ratio based on the nucleon coalescence model
are given in Section \ref{cluster}. Finally, we conclude with a summary in
Section \ref{summary}.

\section{Momentum dependence of nuclear mean-field potential}

\label{momentum}

The energy per nucleon in an asymmetric nuclear matter is usually expressed
as 
\begin{equation}
E(\rho ,\delta )=E(\rho ,\delta =0)+E_{\text{\textrm{sym}}}(\rho )\delta
^{2}+\mathcal{O}(\delta ^{4}),  \label{eos}
\end{equation}%
where $\rho =\rho _{n}+\rho _{p}$ is the baryon density with $\rho _{n}$ and 
$\rho _{p}$ denoting the neutron and proton densities, respectively; $\delta
=(\rho _{n}-\rho _{p})/(\rho _{p}+\rho _{n})$ is the isospin asymmetry; and $%
E(\rho ,\delta =0)$ is the energy per particle in a symmetric nuclear
matter, while $E_{\mathrm{sym}}(\rho )$ is the nuclear symmetry energy.
Studies based on various many-body theories using non-local interactions
have shown that the momentum dependence of nuclear single-particle potential
is different for neutrons and protons in asymmetric nuclear matter, see
e.g., Ref. \cite{bom} for a review. Using the Gogny effective interaction
(MDI), the nucleon single-particle potential was recently determined in the
mean-field approximation by fitting the saturation properties of nuclear
matter at zero temperature with compressibility $K_{0}=211$ MeV and symmetry
energy 30 MeV \cite{das03}. The resulting potential $U(\rho ,\delta ,\mathbf{%
p},\tau )$ for a nucleon with isospin $\tau $ (1/2 for protons and -1/2 for
neutrons) and momentum $\mathbf{p}$ in an asymmetric nuclear matter with
isospin asymmetry $\delta $ and density $\rho $, can be parameterized as 
\begin{eqnarray}
U(\rho ,\delta ,\mathbf{p},\tau ) &=&A_{u}\frac{\rho _{\tau ^{\prime }}}{%
\rho _{0}}+A_{l}\frac{\rho _{\tau }}{\rho _{0}}+B\left( \frac{\rho }{\rho
_{0}}\right) ^{\sigma }(1-x\delta ^{2})-8\tau x\frac{B}{\sigma +1}\frac{\rho
^{\sigma -1}}{\rho _{0}^{\sigma }}\delta \rho _{\tau ^{\prime }}  \notag \\
&+&\frac{2C_{\tau ,\tau }}{\rho _{0}}\int d^{3}\mathbf{p}^{\prime }\frac{%
f_{\tau }(\mathbf{r},\mathbf{p}^{\prime })}{1+(\mathbf{p}-\mathbf{p}^{\prime
})^{2}/\Lambda ^{2}}+\frac{2C_{\tau ,\tau ^{\prime }}}{\rho _{0}}\int d^{3}%
\mathbf{p}^{\prime }\frac{f_{\tau ^{\prime }}(\mathbf{r},\mathbf{p}^{\prime
})}{1+(\mathbf{p}-\mathbf{p}^{\prime })^{2}/\Lambda ^{2}},  \label{mdi}
\end{eqnarray}%
where $\tau \neq \tau ^{\prime }$ and $f_{\tau }(\mathbf{r},\mathbf{p})$
denotes the phase-space distribution function at coordinate $\mathbf{r}$ and
momentum $\mathbf{p}$. The parameter $x$ in Eq.(\ref{mdi}) is introduced to
reflect the uncertainty of our knowledge on the density dependence of
nuclear symmetry energy, and it has a value of $1$ in the Cogny interaction %
\cite{das03}. For the momentum-independent part of the potential, given by
the first line of Eq.(\ref{mdi}), the value for $\sigma $ is $3/4$ and for $B
$ is $106.35$ MeV, while those for $A_{l}$ and $A_{u}$ depend on the value
of $x$. For $x=0$, we have $A_{l}=-120.57$ MeV and $A_{u}=-95.98$ MeV. For
other values of $x$, they are 
\begin{equation}
A_{l}(x)=A_{l}(0)+\frac{2B}{\sigma +1}x,\qquad A_{u}(x)=A_{u}(0)-\frac{2B}{%
\sigma +1}x,
\end{equation}%
with $A_{l}(0)$ and $A_{u}(0)$ denoting the values for $x=0$.

The terms with parameters $C_{unlike}=-103.4$ MeV and $C_{like}=-11.7$ MeV
with $\Lambda =1.0p_{F}^{0}$, where $p_{F}^{0}$ denotes the Fermi momentum
at normal nuclear matter density, in the second line of Eq.(\ref{mdi})
describes the momentum dependence of not only the nuclear isoscalar
potential but also the nuclear symmetry potential as a nucleon with isospin $%
\tau $ interacts differently with unlike and like nucleons in the background
fields.

\begin{figure}[th]
\includegraphics[scale=1.2]{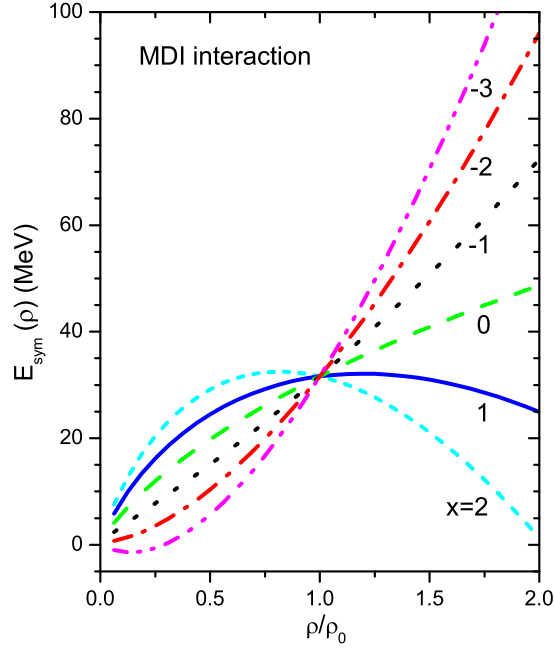}
\caption{{\protect\small (Color online) Density dependence of nuclear
symmetry energy from the MDI interaction with different }$x${\protect\small %
\ values.}}
\label{esym}
\end{figure}

With different $x$ values, one can obtain different density dependence for
the nuclear symmetry energy without changing other properties of the
asymmetric nuclear matter. As seen in Fig. \ref{esym}, where the density
dependence of nuclear symmetry energy from the MDI interaction with
different $x$ values are shown, the symmetry energy becomes stiffer with
decreasing $x$ but its value at normal nuclear density remains the same. To
explore the large range of $E_{sym}(\rho )$ predicted by various many-body
theories \cite{brown,wiringa,bom1,greco01}, we consider in the present study
two values of $x=1$ and $x=-2$. Since the nuclear symmetry energy given by $%
x=1$ has a weaker dependence on density than the one given by $x=-2$, we
call it a \textsl{soft} symmetry energy while the one corresponding to $x=-2$
is called a \textsl{hard} symmetry energy. We note that the soft symmetry
energy with $x=1$ is the same as that given by the default Cogny interaction.

\begin{figure}[th]
\includegraphics[scale=1.2]{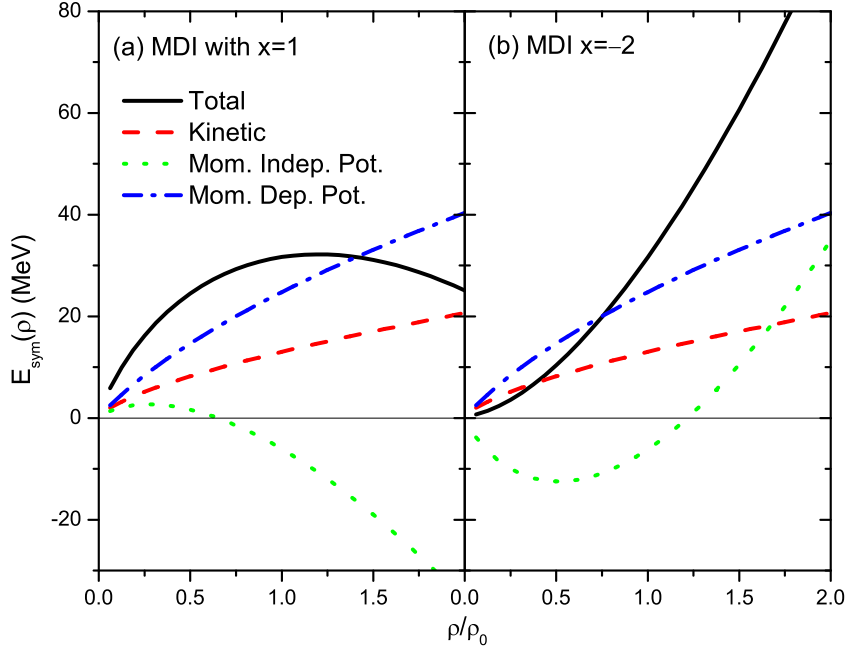}
\caption{{\protect\small (Color online) Density dependence of total symmetry
energy together with contributions from the kinetic energy part as well as
the momentum-independent and momentum-dependent potential energy parts from
the MDI interaction with (a) }$x=1${\protect\small \ and (b) }$x=-2$%
{\protect\small .}}
\label{esymSH}
\end{figure}

The difference in the density dependence of soft and stiff symmetry energies
is due to the different density dependence in the momentum-independent part
of the potential from the MDI interaction as shown in Fig. \ref{esymSH},
where we give the density dependence of total symmetry energy for these two
cases together with that due to contributions from the kinetic energy part
as well as the momentum-independent and momentum-dependent potential energy
parts. The total potential energy contribution to the symmetry energy shown
in Fig.\ref{esymSH} can be parameterized as 
\begin{equation}
E_{sym}^{pot}(\rho )=3.08+39.6u-29.2u^{2}+5.68u^{3}-0.523u^{4}\text{ (MeV) },%
\text{ }  \label{mdi1}
\end{equation}%
for the soft symmetry energy, i.e., the MDI with $x=1$ and 
\begin{equation}
E_{sym}^{pot}(\rho )=-1.83-5.45u+30.34u^{2}-5.04u^{3}+0.45u^{4}\text{ (MeV) }
\label{mdim2}
\end{equation}%
for the hard symmetry energy, i.e., the MDI with $x=-2$. In the above, $%
u\equiv \rho /\rho _{0}$ is the reduced nucleon density.

To study the effects due to the momentum dependence of nuclear symmetry
energy, we also consider a nucleon potential $U_{noms}(\rho ,\delta ,\mathbf{%
p},\tau )\equiv U_{0}(\rho ,\mathbf{p})+U_{sym}(\rho ,\delta ,\tau )$ with
the isoscalar part taken from the original MDYI interaction \cite{gale90},
i.e., 
\begin{equation}
U_{0}(\rho ,\mathbf{p})=-110.44u+140.9u^{1.24}-\frac{130}{\rho _{0}}\int
d^{3}\mathbf{p}^{\prime }\frac{f(\mathbf{r},\mathbf{p}^{\prime })}{1+(%
\mathbf{p}-\mathbf{p}^{\prime })^{2}/(1.58p_{F}^{0})^{2}},
\end{equation}%
which has a compressibility $K_{0}=215$ MeV and is almost the same as the
momentum-dependent isoscalar potential given by the MDI interaction, i.e.,
Eq. (\ref{mdi}). For the momentum-independent symmetry potential $%
U_{sym}(\rho ,\delta ,\tau )$, it is obtained from $U_{sym}(\rho ,\delta
,\tau )=\partial W_{sym}/\partial \rho _{\tau }$ using the isospin-dependent
part of the potential energy density $W_{sym}=E_{sym}^{pot}(\rho )\cdot \rho
\cdot \delta ^{2}$ where $E_{sym}^{pot}(\rho )$ is given by Eqs (\ref{mdi1})
and (\ref{mdim2}) from the MDI interaction.

To study the effects due to the momentum dependence of isoscalar nuclear
potential, we also consider the usual momentum-independent soft nuclear
isoscalar potential with $K_{0}=200$ MeV (SBKD), firstly introduced by
Bertsch, Kruse and Das Gupta\cite{BKD84}, i.e., 
\begin{equation}
U(\rho )=-356~u+303~u^{7/6}.
\end{equation}

\begin{figure}[th]
\includegraphics[scale=1.5]{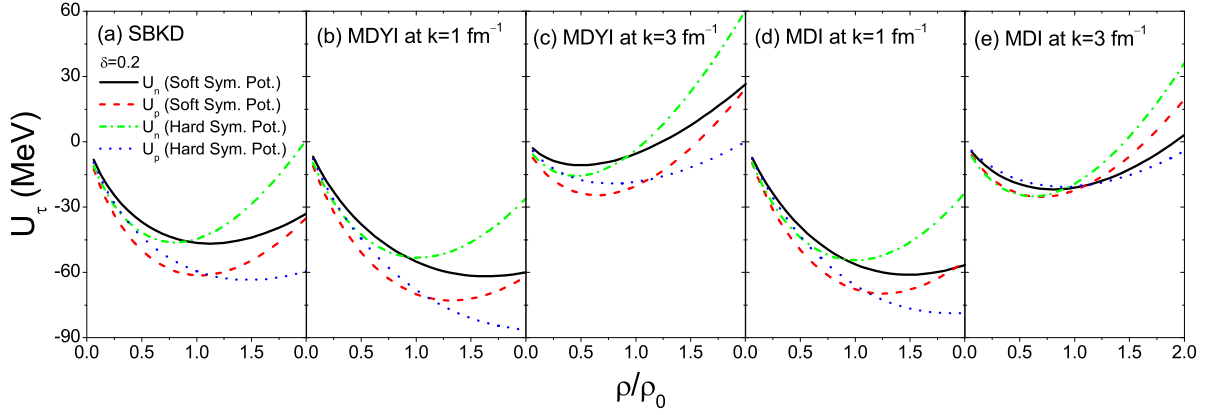}
\caption{{\protect\small (Color online) Density dependence of neutron and
proton single-particle potentials at different momenta in asymmetric nuclear
matter with isospin asymmetry $\protect\delta=0.2$ for different nucleon
effective interactions.}}
\label{UnpRho}
\end{figure}

The density dependence of neutron and proton single-particle potentials at
isospin asymmetry $\delta =0.2$ and at momenta $k=1$ fm$^{-1}$ and $k=3$ fm$%
^{-1}$ are shown in Fig. \ref{UnpRho} for the SBKD, MDYI, and MDI
interactions. It is seen that neutrons generally have a stronger repulsive
potential than protons. Their difference further depends on the nuclear
symmetry energy, with the soft one giving a larger difference at low
densities while the hard one giving a larger difference at high densities.
Also, with momentum-dependent MDYI and MDI interactions, the nuclear
potential is more repulsive for high momentum nucleons. Moreover, the
momentum dependence of nuclear symmetry potential from the MDI interaction
generally reduces the difference between the neutron and proton potentials,
especially at high momenta.

\begin{figure}[th]
\includegraphics[scale=1.4]{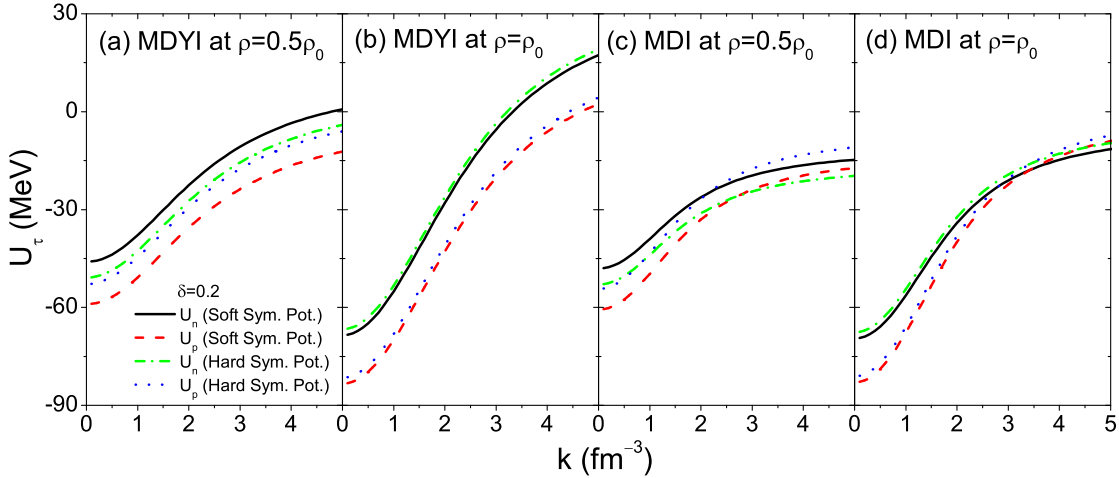}
\caption{{\protect\small (Color online) Momentum dependence of neutron and
proton single-particle potentials in asymmetric nuclear matter with isospin
asymmetry $\protect\delta=0.2$ and at different densities for different
nucleon effective interactions.}}
\label{UnpK}
\end{figure}

To see more clearly the momentum dependence of nucleon mean-field potential,
we show in Fig. \ref{UnpK} the momentum dependence of neutron and proton
single-particle potentials in asymmetric nuclear matter with isospin
asymmetry $\delta =0.2$ and at different densities for different nucleon
effective interactions. Although the MDI interaction is seen to give a
similar momentum dependence in the nucleon potential as the MDYI
interaction, the difference between the neutron and proton potentials, is
reduced, especially at higher momenta, due to its momentum dependence in the
nuclear symmetry potential. Therefore, the MDI interaction gives a weaker
symmetry potential than that from the MDYI interaction.

\begin{figure}[th]
\includegraphics[scale=1.1]{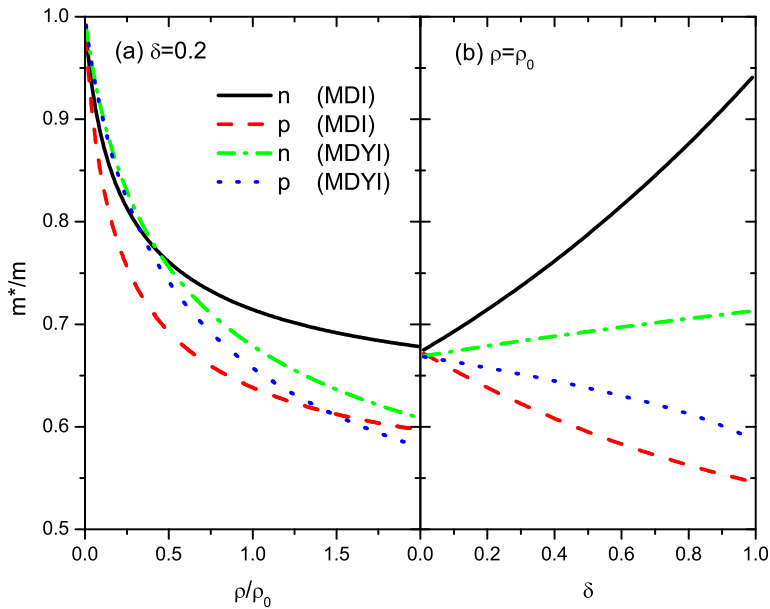}
\caption{{\protect\small (Color online) (a) Density and (b) isospin
asymmetry dependence of nucleon effective mass in asymmetric nuclear matter
with isospin asymmetry }$\protect\delta =0.2$ {\protect\small and at normal
nuclear matter density for the MDYI and MDI interactions.}}
\label{mstar}
\end{figure}

The nucleon effective mass is an important quantity to characterize the
momentum dependence of nuclear mean-field potential. From the
single-particle potential, we can obtain the nucleon effective mass from 
\begin{equation}
\frac{m_{\tau }^{\ast }}{m_{\tau }}=\left( 1+\frac{m_{\tau }}{p}\frac{dU}{dp}%
\right) _{p=p_{f}(\tau )}^{-1}.
\end{equation}%
In Fig. \ref{mstar}, we show both the density (a) and isospin asymmetry (b)
dependence of nucleon effective mass in asymmetric nuclear matter with
isospin asymmetry $\delta =0.2$ and at normal nuclear matter density for the
MDYI and MDI interactions. It is seen that a neutron has a larger effective
mass than a proton in neutron-rich nuclear matter and both depends strongly
on the isospin asymmetry of nuclear matter. The small difference between the
neutron and proton effective masses for the MDYI interaction is due to the
different neutron and proton Fermi energies. In addition, the MDYI
interaction shows a slightly stronger density dependence for the nucleon
effective mass than the MDI interaction.

Above discussions indicate that the nucleon symmetry potential is reduced
after including its momentum dependence and the effect gets larger with
increasing nucleon momentum. The difference between symmetry potentials with
and without momentum dependence is thus larger for nucleons with higher
momenta. In the following, the three potentials obtained with the SBKD,
MDYI, and MDI interactions will be used together with the isospin-dependent
experimental NN cross sections in the IBUU model \cite{li97,li04a,li04b} to
study the effects due to the momentum dependence of isoscalar and isovector
(symmetry) potentials on two-nucleon correlation functions and light cluster
production in intermediate energy heavy-ion collisions induced by
neutron-rich nuclei. In previous studies \cite{ireview98,li97,CorShort}, it
has been shown that these observables are sensitive to the symmetry
potential but not to the isoscalar potential and the in-medium
nucleon-nucleon cross sections.

\section{Two-nucleon correlation functions}

\label{correlation}

The space-time properties of nucleon emission source, which are important
for understanding the reaction dynamics of heavy-ion collisions, can be
extracted from two-particle correlation functions; see, e.g., Refs. \cite%
{Boal90,bauer,ardo97,Wied99} for reviews. In most studies, only the
two-proton correlation function is studied \cite%
{gong90,gong91,gong93,kunde93,handzy95,Verde02,Verde03}. Recently, data on
two-neutron and neutron-proton correlation functions have also become
available. The neutron-proton correlation function is especially useful as
it is free of correlations due to wave-function anti-symmetrization and
Coulomb interactions. Indeed, Ghetti \textit{et al.} have deduced from
measured neutron-proton correlation function the emission sequence of
neutrons and protons in intermediate energy heavy-ion collisions \cite%
{Ghetti00,Ghetti01,Ghetti03} and have also studied the isospin effects on
two-nucleon correlation functions \cite{Ghetti04}.

In standard Koonin-Pratt formalism \cite{koonin77,pratt1,pratt2}, the
two-particle correlation function is obtained by convoluting the emission
function $g(\mathbf{p},x)$, i.e., the probability for emitting a particle
with momentum $\mathbf{p}$ from the space-time point $x=(\mathbf{r},t)$,
with the relative wave function of the two particles, i.e., 
\begin{equation}
C(\mathbf{P},\mathbf{q})=\frac{\int d^{4}x_{1}d^{4}x_{2}g(\mathbf{P}%
/2,x_{1})g(\mathbf{P}/2,x_{2})\left| \phi (\mathbf{q},\mathbf{r})\right| ^{2}%
}{\int d^{4}x_{1}g(\mathbf{P}/2,x_{1})\int d^{4}x_{2}g(\mathbf{P}/2,x_{2})}.
\label{CF}
\end{equation}%
In the above, $\mathbf{P(=\mathbf{p}_{1}+\mathbf{p}_{2})}$ and $\mathbf{q(=}%
\frac{1}{2}(\mathbf{\mathbf{p}_{1}-\mathbf{p}_{2}))}$ are, respectively, the
total and relative momenta of the particle pair; and $\phi (\mathbf{q},%
\mathbf{r})$ is the relative two-particle wave function with $\mathbf{r}$
being their relative position, i.e., $\mathbf{r=(r}_{2}\mathbf{-r}_{1}%
\mathbf{)-}$ $\frac{1}{2}(\mathbf{\mathbf{v}_{1}+\mathbf{v}_{2})(}t_{2}-t_{1}%
\mathbf{)}$. This approach has been very useful in studying effects of
nuclear equation of state and nucleon-nucleon cross sections on the reaction
dynamics of intermediate energy heavy-ion collisions \cite{bauer}. In the
present paper, we use the Koonin-Pratt method to determine the
nucleon-nucleon correlation functions in order to study the effect due to
the momentum dependence of nuclear mean-field potential and the density
dependence of nuclear symmetry energy on the spatial and temporal properties
of nucleon emission source in intermediate energy heavy-ion collisions.

\begin{figure}[th]
\includegraphics[scale=1.4]{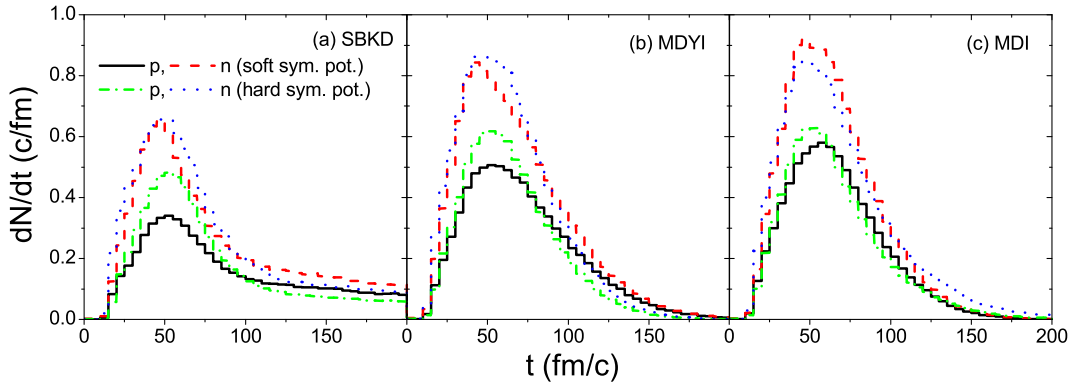}
\caption{{\protect\small (Color online) Emission rates of protons and
neutrons as functions of time for different nucleon effective interactions.}}
\label{emRate}
\end{figure}

As an example, we study here central collisions of $^{52}$Ca + $^{48}$Ca at $%
E=80$ \textrm{MeV/nucleon}. This particular reaction system with isospin
asymmetry $\delta =0.2$ can be studied at the future Rare Isotope
Accelerator (RIA). In the present work, nucleons are considered as emitted
when their local densities are less than $\rho _{0}/8$ and subsequent
interactions do not cause their recapture into regions of higher density. In
Fig. \ref{emRate}, we show the emission rates of protons and neutrons as
functions of time for the SBKD, MDYI, and MDI interactions with soft and
hard symmetry energies. It is clearly seen that there are two stages of
nucleon emissions; an early fast emission and a subsequent slow emission.
This is consistent with the long-lived nucleon emission source observed in
previous BUU calculations \cite{handzy95}. For the momentum-independent
nuclear potential (SBKD), we find from Fig. \ref{emRate} (a) that the hard
symmetry energy enhances the emission of early high momentum protons
(dash-dotted line) and neutrons (dotted line) but suppresses late slow
emission compared with results from the soft symmetry energy (protons and
neutrons are given by solid and dashed lines, respectively). The difference
between the emission rates of protons and neutrons is, however, larger for
the soft symmetry energy. Fig. \ref{emRate} (b) shows results from the MDYI
interaction which includes the momentum-dependent isoscalar potential but
the momentum-independent symmetry potential. It is seen that the momentum
dependence of isoscalar potential enhances significantly the nucleon
emission rate due to the more repulsive momentum-dependent nuclear potential
at high momenta. As a result, the relative effect due to the symmetry
potential is reduced compared with the results shown in Fig. \ref{emRate}
(a). In Fig. \ref{emRate} (c), we use the MDI interaction which includes
momentum dependence in both isoscalar potential and symmetry potential. The
momentum dependence of symmetry potential leads to a slightly faster nucleon
emission but the symmetry potential effects are reduced. We note that the
fraction of total number of emitted nucleons, i.e., before $200$ fm/c in the
IBUU simulations, is about $80\%$ for the SBKD interaction but almost $100\%$
for the MDYI and MDI interactions. The enhancement of nucleon emissions with
momentum-dependent nuclear mean-field potential was also observed in
previous calculations \cite{greco99}. It should be noted that the emitted
nucleons in the present study are not exactly those observed experimentally
since we have not included explicitly the production of clusters and
intermediate mass fragments in the IBUU simulations.

\begin{figure}[th]
\includegraphics[scale=1.1]{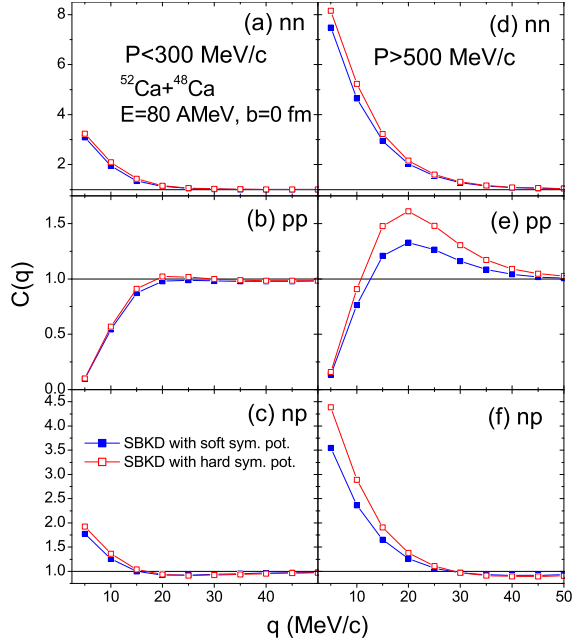}
\caption{{\protect\small (Color online) Two-nucleon correlation functions
gated on the total momentum $P$ of nucleon pairs using the SBKD interaction
with soft (filled squares) or stiff (open squares) symmetry energy. Left
panels are for $P<300$ MeV/c while right panels are for $P>500$ MeV/c.}}
\label{CF35hSBKD}
\end{figure}

Using the program Correlation After Burner \cite{hbt}, which takes into
account final-state nucleon-nucleon interactions, we have evaluated
two-nucleon correlation functions from the emission function given by the
IBUU model. Shown in Fig. \ref{CF35hSBKD} are two-nucleon correlation
functions gated on the total momentum $P$ of nucleon pairs from central
collisions of $^{52}$Ca + $^{48}$Ca at $E=80$ \textrm{MeV/nucleon} by using
the SBKD interaction with the soft and hard symmetry potentials. The left
and right panels are for $P<300$ \textrm{MeV/c} and $P>500$ \textrm{MeV/c},
respectively. Both neutron-neutron (upper panels) and neutron-proton (lower
panels) correlation functions peak at $q\approx 0$ \textrm{MeV/c}, while the
proton-proton correlation function (middle panel) is peaked at about $q=20$ 
\textrm{MeV/c} due to the strong final-state s-wave attraction. The latter
is suppressed at $q=0$ as a result of Coulomb repulsion and
anti-symmetrization of the two-proton wave function. These general features
are consistent with those observed in experimental data from heavy-ion
collisions \cite{Ghetti00}.

For nucleon pairs with high total momentum, their correlation function is
stronger for the hard symmetry energy than for the soft symmetry energy:
about $24\%$ and $9\%$ for neutron-proton pairs and neutron-neutron pairs at
low relative momentum $q=5$ MeV/c, respectively, and $21\%$ for
proton-proton pairs at $q=20$ \textrm{MeV/c}. The neutron-proton correlation
function thus exhibits the highest sensitivity to the density dependence in
nuclear symmetry energy $E_{\mathrm{sym}}(\rho )$. For nucleon pairs with
low total momenta, the symmetry potential effects are weak. These results
are again consistent with previous calculations \cite{CorShort,CorLong}.

\begin{figure}[th]
\includegraphics[scale=1.1]{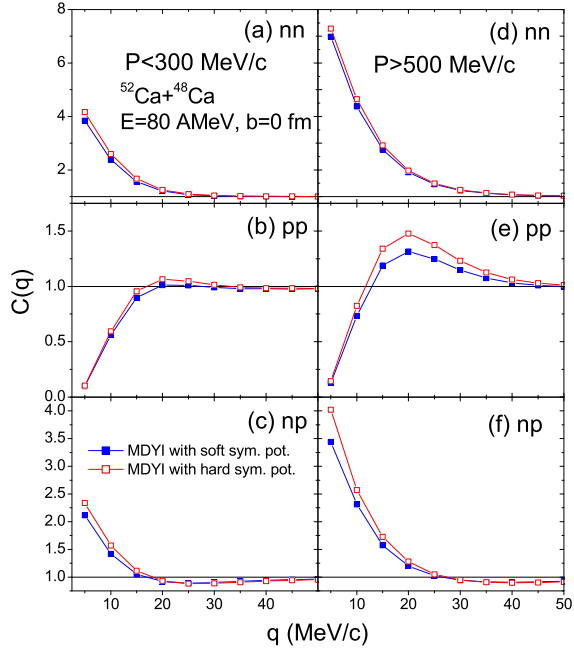}
\caption{{\protect\small (Color online) Same as in Fig. \ref{CF35hSBKD} but
using the MDYI interaction with soft (filled squares) or stiff (open
squares) symmetry energy.}}
\label{CF35hMDYI}
\end{figure}

What will happen to the symmetry energy effect on two-nucleon correlation
functions if we include the momentum-dependent isoscalar potential in the
IBUU model? This can bee seen from Fig. \ref{CF35hMDYI}, where they are
shown for the same central collisions of $^{52}$Ca+$^{48}$Ca at $E=80$ 
\textrm{MeV/nucleon} by using the MDYI interaction with soft and hard
symmetry potentials in the IBUU model. For nucleon pairs with low total
momentum, shown in left panels, their correlation functions remain
insensitive to the nuclear symmetry energy. For nucleon pairs with high
total momentum, shown in right panels, their correlation function is again
stronger for the hard symmetry energy than for the soft symmetry energy:
about $17\%$ and $4\% $ for neutron-proton pairs and neutron-neutron pairs
at low relative momentum $q=5$ MeV/c, respectively, and $12\%$ for
proton-proton pairs at $q=20$ \textrm{MeV/c}. Compared to the results from
the SBKD interaction, the momentum dependence of isoscalar potential thus
reduces the symmetry potential effects on two-nucleon correlation functions.
This is due to the fact that the repulsive momentum-dependent potential
enhances nucleon emissions and thus reduces the density effect on nucleon
emissions, leading to a weaker symmetry potential effects on two-nucleon
correlation functions.

\begin{figure}[th]
\includegraphics[scale=1.1]{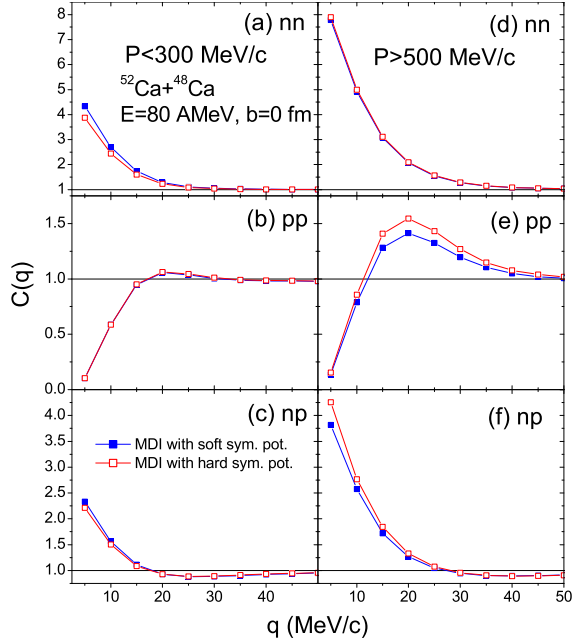}
\caption{{\protect\small (Color online)  Same as in Fig. \ref{CF35hSBKD} but
using the MDI interaction with soft (filled squares) or stiff (open squares)
symmetry energy.}}
\label{CF35hMDI}
\end{figure}

To see how the momentum dependence of nuclear symmetry energy affects the
effect due to its density dependence on two nucleon correlation functions,
we show in Fig.\ref{CF35hMDI} the results from the MDI interaction with the
soft and hard symmetry potentials for central collisions of $^{52}$Ca + $%
^{48}$Ca at $E=80$ \textrm{MeV/nucleon}. It is seen that the effect due to
the stiffness of nuclear symmetry energy is again small for nucleon pairs
with low total momentum as shown in left panels. For nucleon pairs with high
total momentum, shown in right panels, their correlation function remains
stronger for the hard symmetry energy than for the soft symmetry energy:
about $11\%$ for neutron-proton pairs at low relative momentum $q=5$ \textrm{%
MeV/c} and $9\%$ for proton-proton pairs at $q=20$ \textrm{MeV/c}. The
symmetry potential effect on the correlation function of neutron-neutron
pairs is, however, very weak. Compared with results from the SBKD and MDYI
interactions, the two-nucleon correlation functions from the MDI interaction
are thus smaller, mainly due to the very small difference between its
neutron and proton potentials, especially for higher momentum nucleons as
shown in Fig. \ref{UnpRho} and Fig. \ref{UnpK}.

\section{The t/$^{3}$He ratio}

\label{cluster}

Light cluster production has been extensively studied in experiments
involving heavy-ion collisions at all energies, e.g., see Ref. \cite%
{Hodgson03} for a recent review. A popular model for describing the
production of light clusters in these collisions is the coalescence model,
e.g., see Ref. \cite{Csernai86} for a theoretical review, which has been
used at both intermediate \cite{Gyu83,Koch90,Indra00} and high energies \cite%
{Mattie95,Nagle96}. In this model, the probability for producing a cluster
is determined by the overlap of its Wigner phase-space density with the
nucleon phase-space distribution at freeze out. Explicitly, the multiplicity
of a $M$-nucleon cluster in a heavy-ion collision is given by \cite{Mattie95}
\begin{equation}
N_{M}=G\int d\mathbf{r}_{i_{1}}d\mathbf{q}_{i_{1}}\cdots d\mathbf{r}%
_{i_{M-1}}d\mathbf{q}_{i_{M-1}}\langle \underset{i_{1}>i_{2}>...>i_{M}}{\sum 
}\rho _{i}^{W}(\mathbf{r}_{i_{1}},\mathbf{q}_{i_{1}}\cdots \mathbf{r}%
_{i_{M-1}},\mathbf{q}_{i_{M-1}})\rangle .
\end{equation}%
In the above, $\mathbf{r}_{i_{1}},\cdots ,\mathbf{r}_{i_{M-1}}$ and $\mathbf{%
q}_{i_{1}},\cdots ,\mathbf{q}_{i_{M-1}}$ are, respectively, the $M-1$
relative coordinates and momenta taken at equal time in the $M$-nucleon rest
frame; $\rho _{i}^{W}$ is the Wigner phase-space density of the $M$-nucleon
cluster; and $\langle \cdots \rangle $ denotes event averaging. The
spin-isospin statistical factor for the cluster is given by $G$, and its
value is $3/8$ for deuteron and $1/3$ for triton or $^{3}$He, with the
latter including the possibility of coalescence of a deuteron with another
nucleon to form a triton or $^{3}$He \cite{Polleri99}. Details about such
calculation can be found in Ref. \cite{ClstLong}.

\begin{figure}[th]
\includegraphics[scale=1.4]{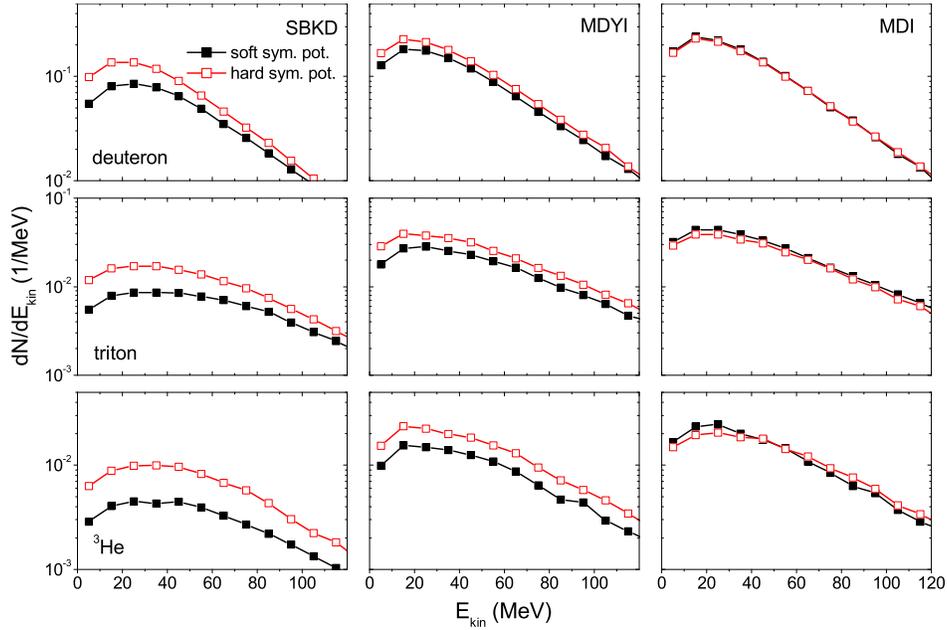}
\caption{{\protect\small (Color online) Kinetic energy spectra in the
center-of-mass system for deuteron (first row), triton (second row), and }$%
^{3}${\protect\small He (third row) from central collisions of }$^{52}$%
{\protect\small Ca + }$^{48}${\protect\small Ca at }$E=80${\protect\small \
MeV/nucleon by using the SBKD (first column), MDYI (second column) and MDI
(third column) interactions with the soft (solid squares) or stiff (open
squares) symmetry energy.}}
\label{dNdE}
\end{figure}

Shown in Fig. \ref{dNdE} are the kinetic energy spectra in the
center-of-mass system for deuterons (first row), tritons (second row), and $%
^{3}$He's (third row) from central collisions of $^{52}$Ca + $^{48}$Ca at $%
E=80$ MeV/nucleon by using the SBKD (first column), MDYI (second column) and
MDI (third column) interactions with the soft (solid squares) and stiff
(open squares) symmetry energies. For the SBKD interaction, the yields of
these light clusters are very sensitive to the density dependence of nuclear
symmetry energy, which is consistent with previous studies \cite%
{ClstShort,ClstLong}. Including momentum dependence in the isoscalar
potential by using the MDYI interaction in the IBUU model reduces slightly
the symmetry potential effect, although the yields of light clusters are
increased significantly as a result of enhanced nucleon emissions from the
momentum-dependent nuclear potential as shown in Fig. \ref{emRate}. If we
further include the momentum dependence in symmetry potential by using the
MDI interaction, the final yields of light clusters become even less
sensitive to the density dependence of symmetry potential than those from
the MDYI interaction. This is due to the very small difference between the
neutron and proton potentials and the momentum dependence of their symmetry
potentials in the MDI interaction as discussed above. It should be noted
that the coalescence model is expected to be a reasonable model for
describing the production of light clusters with large kinetic energies as
they are not contaminated by contributions from decays of heavy fragments,
which are mainly of low kinetic energies and may not be negligible in
intermediate energy heavy-ion collisions.

\begin{figure}[th]
\includegraphics[scale=1.4]{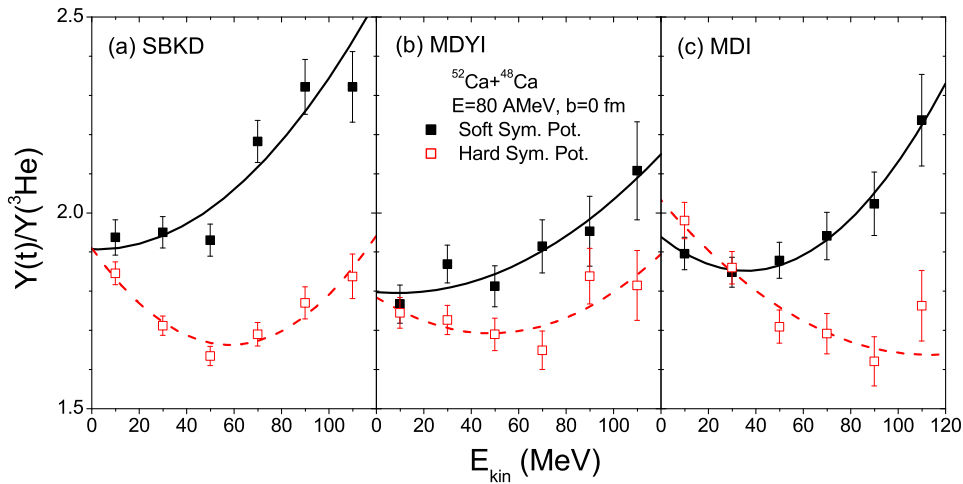}
\caption{{\protect\small (Color online) The t/}$^{3}${\protect\small He
ratio as a function of cluster kinetic energy in the center-of-mass system
for different interactions (a) SBKD, (b) MDYI, and (c) MDI with the soft
(solid squares) and stiff (open squares) symmetry energies. The lines are
drawn to guide the eyes.}}
\label{RtHe3}
\end{figure}

The isobaric yield ratio t/$^{3}$He is less model-dependent and also less
affected by other effects, such as the feedback from heavy fragment
evaporation and the feed-down from produced excited triton and $^{3}$He
states. In Fig. \ref{RtHe3}, we show the t/$^{3}$He ratio with statistical
errors as a function of cluster kinetic energy in the center-of-mass system
for the SBKD, MDYI and MDI interactions with the soft (solid squares) and
stiff (open squares) symmetry energies. For all nuclear potentials, it is
seen that the ratio t/$^{3}$He obtained with different symmetry energies
exhibits very different energy dependence. While the t/$^{3}$He ratio
increases with kinetic energy for the soft symmetry energy, it decreases
and/or increase weakly with kinetic energy for the stiff symmetry energy.
For both soft and stiff symmetry energies, the ratio t/$^{3}$He is larger
than the neutron to proton ratio of the whole reaction system, i.e., \textsl{%
N/Z}$=1.5$. This is in agreement with results from both experiments and the
statistical model simulations for other reaction systems and incident
energies \cite{Hagel00,Cibor00,Sobotka01,Vesel01,Chomaz99}. It is
interesting to note that although the yield of light clusters is not so
sensitive to the density dependence of symmetry potential for the MDI
interaction, the t/$^{3}$He ratio shows very different energy dependence for
the soft and hard symmetry potentials. This is related to the different
momentum dependence of symmetry potential in the MDI interaction, especially
at low densities, as shown in Fig. \ref{UnpK}.

\section{Summary}

\label{summary}

Through studying two-nucleon correlation functions and light cluster
production using an isospin- and momentum-dependent transport model, we have
investigated the effects due to the momentum dependence of isoscalar nuclear
potential and also the symmetry potential on the space-time properties of
nucleon emission source in heavy-ion collisions induced by neutron-rich
nuclei at intermediate energies. It is found that the momentum dependence of
both isoscalar nuclear potential and the symmetry potential influences
significantly the space-time properties of nucleon emission source, leading
thus to appreciable effects on two-nucleon correlation functions and light
cluster production in these collisions. Specifically, the momentum
dependence of nuclear potential reduces the sensitivity of two-nucleon
correlation functions and the light cluster yield on the stiffness of
nuclear symmetry energy. However, the t/$^{3}$He ratio is still found to be
sensitive to the stiffness of symmetry energy after including the momentum
dependence of nuclear potential. The study of these observables in
intermediate energy heavy-ion collisions thus offers the possibility to
probe both the momentum dependence of nuclear symmetry potential and the
density dependence of nuclear symmetry energy.

\begin{acknowledgments}
This paper was based on work supported by the U.S. National Science
Foundation under Grant Nos. PHY-0098805, PHY-0088934 and PHY-0243571 as well
as the Welch Foundation under Grant No. A-1358. LWC was also supported by
the National Natural Science Foundation of China under Grant No. 10105008.
\end{acknowledgments}

\end{document}